\documentclass[12pt]{article}
\usepackage{amssymb}
\usepackage{amsmath}
\usepackage{epsfig}
\usepackage{float}
\newcommand{\be}{\begin{equation}}
\newcommand{\ee}{\end{equation}}
\newcommand{\bea}{\begin{eqnarray}}
\newcommand{\eea}{\end{eqnarray}}
\begin{document}

\begin{center}
{\bf Recoilless resonant neutrino experiment and origin of neutrino
oscillations}\footnote{The report at the  Workshop on Next
Generation Nucleon Decay and Neutrino Detectors, NNN06, September
21-23, 2006, University of Washington, Seattle, USA.}

\end{center}
\begin{center}
S. M. Bilenky
\end{center}

\begin{center}
{\em  Joint Institute for Nuclear Research, Dubna, R-141980,
Russia\\}
\end{center}
\begin{center}
F. von  Feilitzsch and W. Potzel
\end{center}
\begin{center}
{\em Physik-Department E15, Technische Universit\"at M\"unchen, D-85748
Garching, Germany}
\end{center}
\begin{abstract}
We demonstrate that an experiment with  recoilless resonant emission
and absorption of tritium $\bar \nu_{e}$ could have an important
impact on our understanding of the origin of neutrino oscillations.
\end{abstract}
\section{Introduction}
Evidence for neutrino oscillations obtained in the Super-Kamiokande
atmospheric \cite{SK}, SNO solar  \cite{SNO}, KamLAND
reactor\cite{Kamland} and other neutrino experiments
\cite{Cl,Gallex,Sage,SKsol,K2K,MINOS} is one of the most important
recent discoveries in particle physics. Small neutrino masses can not
be generatad by the Standard Higgs mechanism. Some new mechanism of
mass generation is necessary.

All existing neutrino oscillation data can be explained by
three-neutrino mixing.\footnote{Recent results of the MiniBooNE
experiment\cite{Miniboone} do not confirm the LSND indication
\cite{LSND} in favor of existence of more than three massive
neutrinos.} In the framework of three-neutrino mixing, from the
analysis of the Super-Kamiokande atmospheric neutrino data the
following ranges for the largest neutrino mass-squared difference
$\Delta m^2_{23}$ and for the mixing angle were obtained
\cite{SK}\footnote{ Neutrino mass-squared difference is determined
as follows: $\Delta m^2_{ik}=m^2_{k}-m^2_{i}.$}
\begin{equation}\label{1}
 1.5\cdot 10^{-3}\leq \Delta m^{2}_{23} \leq 3.4\cdot
10^{-3}\rm{eV}^{2};~~\sin^{2}2\theta_{23}>0.92.
\end{equation}
From a global analysis of the KamLAND and solar neutrino data it was found \cite{Kamland}
\begin{equation}\label{2}
\Delta m^{2}_{12} =
7.9^{+0.6}_{-0.5}~10^{-5}~\rm{eV}^{2};~~\tan^{2}\theta_{12}=0.40^{+0.10}_{-0.07}.
\end{equation}
Only an upper bound for the angle $\theta_{13}$ is known at present
\cite{Chooz}:
\begin{equation}\label{2a}
\sin^{2}\theta_{13}\leq 5\cdot 10^{-2}.
\end{equation}
In future neutrino oscillation experiments the accuracies of the
neutrino oscillation parameters are planned to be
improved. One of the major aims of future experiments is to determine
the value of the parameter $\theta_{13}$  which is crucial for the
measurement of the $CP$ violation in the lepton sector and the
determination of the character of the neutrino mass spectrum.

From the point of view of field theory, the phenomenon of neutrino
oscillations is based on the following assumptions:
\begin{enumerate}
\item
Neutrino interactions are the SM charged current (CC) and neutral current (NC)
interactions. The leptonic CC and neutrino NC are given by
\begin{equation}\label{3}
j^{\mathrm{CC}}_{\alpha}(x) =2\, \sum_{l=e,\mu,\tau} \bar
\nu_{lL}(x) \gamma_{\alpha}l_{L}(x) ;~~~ j
^{\mathrm{NC}}_{\alpha}(x) =\sum_{l=e,\mu,\tau} \bar \nu_{lL}(x)
\gamma_{\alpha}\nu_{lL}(x).
\end{equation}
\item
The fields of neutrinos with definite masses enter into CC and NC in
{\em the mixed form}
\begin{equation}\label{4}
\nu_{l L}(x)=\,\sum^{3}_{k=1}U_{l k}\,\nu_{k L}(x).
\end{equation}
Here $\nu_{k }(x)$ is the field of neutrino with mass $m_{k}$ and
$U$ is the unitary PMNS matrix \cite{BP,MNS}.
\end{enumerate}
Due to neutrino mixing, Eq.(\ref{4}), the flavor lepton numbers
$L_{e}$, $L_{\mu}$ and $L_{\tau}$ are not conserved in neutrino
transitions. The standard probability of the transition $\nu_{l} \to
\nu_{l'} $ is given by (see \cite{BGG})
\begin{equation}\label{5}
P(\nu_{l} \to\nu_{l'}) = |\sum^{3}_{k=1} U_{l' k} \,e^{-i \Delta
m^2_{ 1k }\frac{L }{2 E }}\, U^{*}_{lk}|^{2},
\end{equation}
where $L$ is the distance between the neutrino-detection and
neutrino-product\-ion points and $E$ is the neutrino energy.
Expression (\ref{5}) perfectly describes the existing neutrino
oscillation data. Let us notice that taking into account the
unitarity of the mixing matrix we can rewrite (\ref{5}) in a  form
\begin{equation}\label{6}
P(\nu_{l} \to\nu_{l'})=|\delta_{l' l}+ \sum_{k \ne 1} U_{l' k}
\,(e^{-i \Delta  m^2_{  1k }\frac{L }{2 E }} -1) U^{*}_{lk}|^{2}.
\end{equation}
The probabilities $P(\nu_{l} \to\nu_{l'})$  depend on six
parameters. However, $\frac{\Delta m^2_{  12 }} {\Delta m^2_{ 23
}}\ll1$ and $\sin^{2}\theta_{13}\ll 1.$ In first approximation
we can neglect the contributions of these parameters to the transition
probabilities. In this approximation neutrino oscillations in the
atmospheric-LBL and solar-KamLAND regions are described by the
simplest two-neutrino expressions which depend, correspondingly,  on
$\Delta m^2_{ 23 },~ \sin^{2}2\theta_{23}$ and $\Delta  m^2_{1 2
},~\tan^{2}\theta_{12}$ (see review \cite{BGG}). The numerical
values of these parameters given in  (\ref{1}) and  (\ref{2}) were
obtained from the analysis of the experimental data by using
two-neutrino expressions.

Several derivations of  Eq. (\ref{5}), based on  different physical
assumptions, exist in the literature. It is not possible to test
these assumptions in usual neutrino oscillation experiments. We will
show here that an experiment on resonant recoilless emission and capture of tritium $\bar
\nu_{e}$'s, proposed recently in
\cite{Raghavan, Potzel}, could provide such a possibility (for additional information, see \cite{BilPhys}).

\section{Different approaches to neutrino oscillations}
We discuss here different points of view on the origin of neutrino
oscillations. Neutrinos are produced in CC weak processes. For the
difference of momenta of neutrinos with masses $m_{k}$ and $m_{i}$
(in the rest-frame of the source) we have
\begin{equation}\label{7}
\Delta p_{ik}= (p_{k}- p_{i}) \sim \frac{\Delta m^{2}_{ik}}{E},
\end{equation}
where $E$ is the neutrino energy. In standard neutrino oscillation
experiments $E\gtrsim$ MeV. From (\ref{7}), (\ref{1}), and (\ref{2})
follows that $|\Delta p_{ik}|$ is much smaller than the
quantum-mechanical uncertainty of the momentum. This is the reason
why in CC neutrino-production processes together with
$e^{+},\mu^{+},\tau^{+} $, correspondingly, flavor neutrinos
$\nu_{e}, \nu_{\mu}, \nu_{\tau}$,  which are described by {\em mixed
flavor neutrino states},
\begin{equation}\label{10}
|\nu_{l}\rangle =\sum^{3}_{k=1}U_{l
k}^*~|\nu_{k}\rangle,~~~l=e,\mu,\tau
\end{equation}
are produced (see, for example, \cite{BilG}). Here $|\nu_{k}\rangle$
is the state of a neutrino with mass $m_{k}$ and 4-momentum
$p_{k}=(E_{k}, \vec{p_{k}})$.

The main difference between different approaches to neutrino
oscillations is connected with assumptions about {\em the
propagation of flavor neutrino states}. We will discuss two
different assumptions.
\newpage
\begin{center}
 {\bf I. Evolution in time.}
\end{center}
The evolution equation of any quantum system is the Schr\"odinger equation
(see, for example, \cite{BogShir})
\begin{equation}
 i\,~ \frac{\partial |\Psi(t)\rangle}{\partial t} =
H\,~ |\Psi(t)\rangle     \,. \label{11}
\end{equation}
Here $|\Psi(t)\rangle$ is the state of the system at the time $t$
and $H$ is the total Hamiltonian. The general solution of this
equation has the form
\begin{equation}\label{12}
|\Psi(t)\rangle = e^{-i\,H t}\,|\Psi(0)\rangle,
\end{equation}
where $|\Psi(0)\rangle $ is the state of the system at the initial
time $t=0$.

If at $t=0$ the flavor neutrino  $\nu_{l}$ is produced, we have
$|\Psi(0)\rangle =|\nu_{l}\rangle $ and the neutrino state in
vacuum at the time $t\geq 0$ is given by
\begin{equation}\label{13}
|\nu_{l}\rangle_{t}=e^{-i\,Ht}\,|\nu_{l}\rangle=
\sum^{3}_{k=1}e^{-iE_{k}t}\,~U_{lk }^*|\nu_{k}\rangle.
\end{equation}
Thus, if the energies $E_{k}$ are different, the neutrino state
$|\nu_{l}\rangle_{t}$ is a \textit{non-stationary} one. For such states
the time-energy uncertainty relation
\begin{equation}\label{14}
\Delta E~ \Delta t \gtrsim 1
\end{equation}
holds (see, for example, \cite{Sakurai}). In this relation,
$\Delta E$ is the energy uncertainty and $\Delta t $ is the time
interval during which the state of the system is significantly
changed.

Neutrinos are detected via the observation of CC and NC reactions. In
such reactions, flavor neutrinos $\nu_{l'}$, which are described by
mixed coherent states (\ref{10}), are detected. From (\ref{10}) and
(\ref{13}) we find
\begin{equation}\label{15}
|\nu_{l}\rangle_{t}=\sum_{l'}|\nu_{l'}\rangle~\sum^{3}_{k=1}
U_{l'k}~e^{-iE_{k}t}\,~U_{lk }^*.
\end{equation}
Thus, the transition probability $\nu_{l}\to \nu_{l'}$ is given by
\begin{equation}\label{16}
P(\nu_{l}\to \nu_{l'}) =|\sum^{3}_{k=1}
U_{l'k}~e^{-i(E_{k}-E_{1})t}\,~U_{lk }^*|^{2}.
\end{equation}
From this expression it is obvious that if the energies of the
neutrinos with different masses are equal, $P(\nu_{l}\to
\nu_{l'})=\delta_{l'l}$. {\em Thus, in the approach based on the
Schr\"odinger evolution equation, there will be no neutrino
oscillations if $E_{k}=E_{1}$} \cite{BilenkyMat}.

Let us assume now that the flavor neutrino states $ |\nu_{l}\rangle
$ are superpositions of the neutrino states $\nu_{k}$ {\em with the
same momentum $\vec{p}$} and different energies. In the case of
ultrarelativistic neutrinos, which we are interested in, we have
\begin{equation}\label{17}
E_{k} =\sqrt{p^{2}+m^{2}_{k}}\simeq p+\frac{m^{2}_{k}}{2E},
\end{equation}
with $E$ being the neutrino energy at $m_{k}\to 0$.

Taking into account that
\begin{equation}\label{18}
t\simeq L
\end{equation}
we obtain from (\ref{16}) the standard expression (\ref{5}) for the
transition probability. Let us note that the time-energy
uncertainty relation (\ref{14}) takes the form of the well-known
condition for the observation of neutrino oscillations (see
\cite{BilPont}):
\begin{equation}\label{19}
(E_{k}-E_{1})~t \simeq \frac{\Delta m^{2}_{1k}}{2E}L\gtrsim 1.
\end{equation}

It is evident that in the approach based on the Schr\"odinger
equation, oscillations between different flavor neutrinos are due to
the fact that the neutrino state $|\nu_{l}\rangle_{t}$ is a
superposition of states with \textit{different energies}.\footnote{
We assumed that the states of flavor neutrinos are superpositions of
states of neutrinos with different masses and the same momentum. Let
us notice that if we assume that neutrinos with different masses
have different momenta, in the expression for the transition
probability in addition to the standard phases $\frac{\Delta
m^{2}_{1k}}{2E}L$ we will have terms $(p_{k}- p_{1})L$. These
additional terms could be of the same order as the standard phases
and could be different in different experiments. All analyses of
neutrino oscillation data  do not favor such a possibility.}

\begin{center}
 {\bf II. Propagation in  space and time}
\end{center}

It has been suggested in several papers (see
\cite{Lipkin,Kayser,Giunti}) that the mixed neutrino state at the
space-time point $x=(t, \vec{x})$ is given by
\begin{equation}\label{20}
|\nu_{l}\rangle_{x}= \sum^{3}_{k=1}e^{-ip_{k}x}\,~U_{lk
}^*~|\nu_{k}\rangle.
\end{equation}
Here  $|\nu_{k}\rangle$ is the state of a neutrino with mass $m_{k}$
and momentum $p_{k}$. From  (\ref{20}) we find
\begin{equation}\label{21}
|\nu_{l}\rangle_{x}=e^{-ip_{1}x}
\sum_{l'}|\nu_{l'}\rangle~\sum^{3}_{k=1}
U_{l'k}~e^{-i(p_{k}-p_{1})x}\,~U_{lk }^*.
\end{equation}
Thus,
\begin{equation}\label{22}
P(\nu_{l}\to \nu_{l'})=|\sum^{3}_{k=1}
U_{l'k}~e^{-i(p_{k}-p_{1})x}\,~U_{lk }^*|^{2}
\end{equation}
is the probability to find the flavor neutrino $\nu_{l'}$ at the point
$x$ in the case that at point $x=0$ the mixed flavor neutrino
$\nu_{l}$ was produced. For the phase difference we have
\begin{equation}\label{23}
(p_{k}-p_{1})~x=(E_{k}-E_{1})~t-(|\vec{p_{k}}|-|\vec{p_{1}}|)~L=
\frac{E^{2}_{k}-E^{2}_{1}}{E_{k}+E_{1}}t-
(|\vec{p_{k}}|-|\vec{p_{1}}|)~L,
\end{equation}
where $\vec{p_{k}}=|\vec{p_{k}}|\vec{k}$ and $\vec{k}~\vec{x}=L$,
$\vec{k}$ being the unit vector in the direction of the neutrino
momenta.

In the framework of the propagation of the flavor states in time and
space two scenarios were considered in the literature.
\begin{center}
\textbf{Scenario I.}~~$p_{k}\neq p_{1} $, $t\simeq L.$
\end{center}
From (\ref{23}) we find for the oscillation-phase difference
\begin{equation}\label{24}
(p_{k}-p_{1})x=\frac{E^{2}_{k}-E^{2}_{1}}{E_{k}+E_{1}}~t-
(|\vec{p_{k}}|-|\vec{p_{1}}|)~L \simeq
(|\vec{p_{k}}|-|\vec{p_{1}}|)(\frac{|\vec{p_{k}}|+|\vec{p_{1}}|}{E_{k}+E_{1}}~t
-L)+\frac{\Delta m^{2}_{1k}}{2E}~t
\end{equation}
If we assume that the distance and the time are connected by the
relation $t\simeq L$, we can neglect the first term in (\ref{24})
and come  to the standard oscillation-phase difference
\begin{equation}\label{25}
(p_{k}-p_{1})~x\simeq \frac{\Delta m^{2}_{1k}}{2E}~L
\end{equation}
and to the standard expression (\ref{5}) for the transition
probability.
\newpage

\begin{center}
\textbf{Scenario II.}~~~~$E_{k}= E_{1} $, stationary states.
\end{center}
It has been suggested in \cite{Stodol,Lipkin,Kayser} that time is
not measured in neutrino oscillation experiments and in order
to oscillate neutrinos with different masses must have the same
energies. In other words it has been suggested in
\cite{Stodol,Lipkin,Kayser} that the flavor neutrino states are
stationary ones. Taking into account that in this case
$|\vec{p_{k}}|=E -\frac{m^{2}_{k}}{2E}$, from (\ref{23}) we obtain
the standard expression for the oscillation-phase difference
\begin{equation}\label{26}
(p_{k}-p_{1})~x=\frac{\Delta m^{2}_{1k}}{2E}~L
\end{equation}
and the standard expression (\ref{5}) for the probability of the
transition between different flavor neutrinos.

Thus, in all three cases which we have considered we came to the same
expression (\ref{5}) for the transition probability. This means that in
usual neutrino oscillation experiments it is impossible to
distinguish these three cases.

Recently, a new type of neutrino experiment,  based on the
M\"ossbauer effect, has been proposed \cite{Raghavan,Potzel}. In the
next section we will discuss this proposal from the point of view of providing
a possibility to distinguish the different assumptions on the
propagation of mixed neutrino states.

\section{Recoilless resonant emission and absorption of tritium $\bar\nu_{e}$'s}

In \cite{Raghavan}, an experiment has been proposed for the detection of
$\bar\nu_{e}$ with energy $\simeq$ 18.6 keV in recoilless resonant (M\"ossbauer)
transitions:
\begin{equation}\label{27}
 ^{3}\rm{H}\to ^{3}\rm{He}+\bar\nu_{e};~~~\bar\nu_{e}+
^{3}\rm{He}\to ^{3}\rm{H}.
\end{equation}
It was estimated in \cite{Raghavan} that the cross section of the
resonant absorption of $\bar\nu_{e}$ by $^{3}\rm{He}$ is equal to
$\sigma_{R} \simeq 3\cdot 10^{-33}\rm{cm}^{2}$.

The study of neutrino oscillations, driven by $\Delta m^{2}_{23}$,
in an experiment with neutrinos produced and detected according to (\ref{27})
was proposed in \cite{Raghavan}. In such an experiment with a
baseline of $\sim$ 10 m the parameter $\sin^{2}\theta_{13}$ can be
measured.

It was estimated in \cite{Raghavan} that the uncertainty in
energy of the produced antineutrinos in the proposed experiment is of
the order
\begin{equation}\label{28}
\Delta E \simeq 8.6 \cdot 10^{-12}~\rm{eV}.
\end{equation}
This value of $\Delta E $ is much smaller than the energy difference of $\nu_{3}$ and $\nu_{2}$
\begin{equation}\label{29}
(E_{3}-E_{2})\sim \frac{\Delta m^{2}_{23}}{2E}\simeq 6.5 \cdot
10^{-8}~\rm{eV},
\end{equation}
which drive neutrino oscillations in  the approach based on
evolution in time.

Thus, the state of flavor $\bar\nu_{e}$ produced and detected in the
reactions (\ref{27}) is the superposition of states of neutrinos
with  {\em the same energy} and (because of the dispersion relation)
different momenta. This means that in the experiment proposed in
\cite{Raghavan} neutrino oscillations will not be observed if the
approach based on the Schr\"odinger evolution equation is correct.
On the other hand, neutrino oscillations can be observed in this
experiment if one of the Scenarios I or II, based on the propagation
in space and time, is correct. Thus, an experiment with recoilless
resonant emission and absorption of tritium  $\bar\nu_{e}$'s could
have an important impact on our understanding of the origin of
neutrino oscillations.

\section{Conclusion}

In conclusion we make the following remarks:

\begin{enumerate}

\item
In an  experiment  proposed in \cite{Raghavan} a positive effect of
neutrino oscillations can be observed only in the case that the
parameter $\sin^{2}\theta_{13}$ is not too small. At present, only
an upper bound of this parameter  is known \cite{Chooz}. A positive
result of the  experiment would allow to determine the parameter
$\sin^{2}\theta_{13}$ and to exclude the approach based on the
evolution of the mixed neutrino states in time (see Section 2I).
However, a negative result of such an experiment could be the
consequence of the smallness of the parameter $\sin^{2}\theta_{13}$.
Thus, in the case of a negative result of the  experiment  definite
conclusions on the fundamentals of neutrino oscillations can be
drawn only if the parameter $\sin^{2}\theta_{13}$ will be measured
in future reactor (DOUBLE CHOOZ \cite{2Chooz}, Daya Bay
\cite{Dayabay}) or accelerator (T2K \cite{T2K}, Nova \cite{Nova})
experiments.

\item
We have considered an experiment, proposed in \cite{Raghavan}, on
the search for neutrino oscillations driven by the 'large' $\Delta
m^{2}_{23}$ and based on recoilless resonant absorption of $\bar
\nu_{e}$. The baseline of this experiment is $\thicksim$ 10 m. In
such an experiment the effect of neutrino oscillations will be small
(if present at all) because the amplitude of the oscillations is
limited by the upper bound of the CHOOZ experiment
($\sin^{2}2\theta_{13}\lesssim 2\cdot 10^{-1}$). The question arises
if a similar oscillation experiment driven by the "small" $\Delta
m^{2}_{12}$ could be performed. Such an experiment would require an
about 30 times larger baseline, i.e., $\thicksim$ 300 m. Because
tritium acts as a point-like source the expected number of neutrino
events in such an experiment will be $\thicksim$ 1000 times smaller
than in the 10 m-baseline experiment.

It was estimated in \cite{Raghavan} that with a 10 m baseline about
$10^{3}$ $\bar\nu_{e}$-captures/day can be expected. Thus, in an
experiment with a baseline of $\thicksim$ 300 m only about 1
capture/day could be observed. If we neglect the small contributions
of the terms proportional to $\sin^{2}\theta_{13}$ we find from eq.(\ref{5})
for the $\bar\nu_{e}$ survival probability the following
expression
\begin{equation}\label{31}
P(\bar\nu_{e}\to \bar\nu_{e}) =1 -\frac{1}{2}\sin^{2}2\theta_{12}
(1-\cos\Delta m^{2}_{12}\frac{L}{2E}).
\end{equation}
Taking into account that the amplitude of the neutrino oscillations
is large in this case (see eq.(\ref{2})),
such an experiment might still be feasible
although we regard the estimate given in \cite{Raghavan}
as rather optimistic \cite{Potzel}.

Let us note that according to eq.(\ref{28}) the uncertainty of the
energy of the antineutrinos emitted without recoil  is much smaller
than
\begin{equation}\label{32}
\frac{\Delta m^{2}_{12}}{2E}\simeq 2.1\cdot 10^{-9}~\rm{eV}.
\end{equation}
Thus, all our arguments given above for the possibilities to
distinguish different approaches to the physics of neutrino
oscillations are applicable also in this case. Notice also that two
detectors of the same kind would allow to record the antineutrinos
at two distances ($\thicksim$ 10 m and $\lesssim$ 300 m).

\item
 We considered here different assumptions on the propagation of
neutrino states (in time or in space and time). Different
assumptions on the propagation of neutrino states give the same
transition probabilities in the case of standard neutrino
oscillation experiments. We have shown that a recently proposed
M\"ossbauer-type neutrino experiment \cite{Raghavan,Potzel} could allow to
distinguish the different fundamental assumptions on the propagation of
neutrinos with definite masses.

\end{enumerate}
 We thank T. Schwetz for fruitful discussions.
 S. Bilenky acknowledges the ILIAS program for the support and the
 TRIUMF Theory group  for the hospitality.


\begin{thebibliography}{99}

\bibitem{SK} Super-Kamiokande Collaboration, Y. Ashie {\em et al.,}
Phys. Rev. Lett. \textbf{93} (2004) 101801;
Phys. Rev. \textbf{D71} (2005) 11205.


\bibitem{SNO}SNO Collaboration, Phys. Rev. Lett.
\textbf{87} (2001) 071301;~\textbf{89} (2002) 011301 ;~\textbf{89}
(2002) 011302;~\textbf{92} (2004) 181301.


\bibitem{Kamland}KamLAND Collaboration,
T.Araki {\em et al.}, Phys. Rev. Lett. \textbf{94} (2005) 081801;~
hep-ex/0406035.


\bibitem{Cl}B. T. Cleveland {\it et al.}, Astrophys. J. {\bf
496} (1998) 505.

\bibitem{Gallex} GALLEX Collaboration, W. Hampel
{\it et al.}, Phys. Lett. {\bf B 447} (1999) 127 ;\, GNO
Collaboration, M. Altmann {\it et al.}, Phys. Lett. {\bf B 490}
(2000) 16 ;\, Nucl. Phys. Proc. Suppl. {\bf 91} (2001) 44; Phys. Lett. \textbf{B 616} (2005) 174.

\bibitem{Sage} SAGE Collaboration,
J. N. Abdurashitov {\it et al.},
 Phys. Rev. {\bf C 60} (1999) 055801; \,Nucl. Phys. B (Proc. Suppl.) {\bf 110}
(2002) 315;~\,Nucl. Phys. B (Proc. Suppl.) {\bf 118} (2003) 39.

\bibitem{SKsol} Super-Kamiokande Collaboration, S.~Fukuda {\it et al.}, Phys. Rev. Lett.
 {\bf 86} (2001) 5651.

\bibitem{K2K} K2K Collaboration, M.H. Ahn {\em et al.},
Phys. Rev. Lett. \textbf{90} (2003) 041801.

\bibitem{MINOS} MINOS Collaboration, D.G. Michael {\it et al.}
arXiv:hep-ex/0607088.

\bibitem{Miniboone} MiniBooNE Collaboration, A.A. Aguilar-Arevalo {et al.}, arXiv: 0704.1500 v1 [hep-ex] 11 Apr 2007.

\bibitem{LSND} LSND Collaboration,
 A. Aguilar {et al.}, Phys. Rev. D {\bf 64} (2001) 112007;
 G. Drexlin, Nucl. Phys. B (Proc. Suppl.) {\bf 118} (2003) 146.

\bibitem{Chooz} CHOOZ Collaboration,
 M. Apollonio {\it et al.}, Phys. Lett. B {\bf 466} (1999) 415;
 M. Apollonio {\it et al.}, Eur. Phys. J. C  {\bf 27} (2003) 331;
 hep-ex/0301017.




\bibitem{BP}
B.~Pontecorvo, J. Exptl. Theoret. Phys. \textbf{33} (1957) 549.
[Sov. Phys. JETP \textbf{6} (1958) 429 ]; J. Exptl. Theoret. Phys.
\textbf {34} (1958) 247 [Sov. Phys. JETP \textbf {7} (1958) 172 ].


\bibitem{MNS} Z. Maki, M. Nakagawa, and S. Sakata, Prog. Theor.
Phys. {\bf 28} (1962) 870.


\bibitem{BGG} S.M.\, Bilenky, C.\, Giunti, and
W.\,Grimus. Prog. Part. Nucl. Phys. {\bf 43} (1999) 1.



\bibitem{Raghavan}R.S. Raghavan, hep-ph/0601079.

\bibitem{Potzel}W. Potzel, Phys. Scr. {\bf T127} (2006) 85.

\bibitem{BilPhys} S.M. Bilenky, F. von Feilitzsch, and W. Potzel, J. Phys. G: Nucl. Part. Phys. {\bf 34} (2007) 987; hep-ph/0611285.

\bibitem{BilG} S.M. Bilenky and C. Giunti,  Int. J. Mod. Phys. {\bf A16} (2001) 3931;
 hep-ph/0102320.


\bibitem{BogShir} N.N. Bogolubov and D.V. Shirkov,
Introduction to the theory of quantized fields (Interscience,
New York, 1980).



\bibitem{Sakurai} J.J. Sakurai, Modern Quantum Mechanics,
(Addison Wesley Publishing Company, New York, 1994).

\bibitem{BilenkyMat} S.M. Bilenky and M.D. Mateev, hep-ph/0604044.


\bibitem{BilPont} S.M. Bilenky and B. Pontecorvo, Phys.
Rep. {\bf 41} (1978) 225 .






\bibitem{Lipkin}H.Lipkin, Phys. Lett.
 \textbf{B579} (2004) 355.

\bibitem{Kayser} B. Kayser, hep-ph/0506165.

\bibitem{Giunti} C. Giunti, hep-ph/0409230.

\bibitem{Stodol}L. Stodolsky, Phys. Rev. \textbf{D58} (1998) 036006.







\bibitem{2Chooz} DOUBLE CHOOZ collaboration, S.A. Dazeley  {\it et al.}
Proceedings of the  7th International Workshop on Neutrino Factories
and Superbeams (NuFact 05), Frascati, Italy, 21-26 Jun 2005;
hep-ex/0510060

\bibitem{Dayabay}Jun Cao, Nucl. Phys. Proc. Suppl. {\bf 156} (2006)
229.

\bibitem{T2K} T2K Collaboration,
Y. Hayato  {\it et al.}, Nucl. Phys. B (Proc. Suppl.) {\bf 143} (2005)
269.

\bibitem{Nova} D.S. Ayres {\it et al.} NOvA collaboration,
hep-ex/0503053


\end{thebibliography}
\end{document}